\documentclass[%
 reprint,
 amsmath,amssymb,
 aps,
 pre,
floatfix,
]{revtex4-1}

\usepackage{graphicx}
\usepackage{dcolumn}
\usepackage{bm}
\usepackage{esdiff}
\usepackage{upgreek}
\newcommand{\norm}[1]{\left\lVert #1 \right\rVert}
\usepackage{xcolor}

\graphicspath{{figure/}}

\begin{document}

\preprint{APS/123-QED}

\title{Formation of Spiral-Coils among Self-Propelled Chains}

\author{Yao-Kuan Wang${}^1$, Chien-Jung Lo${}^1$, and Wei-Chang Lo${}^2$}
\email{namnatur@gmail.com}

\affiliation{${}^1$Department of Physics and Graduate Institute of Biophysics, National Central University, Jhongli 32001, Taiwan ROC\\
 ${}^2$Department of Physics, Duke University, Durham, NC 27708, USA}


\date{\today}

\begin{abstract}
We study the dynamics of self-propelled chains with the excluded volume interaction via the Brownian dynamics simulation, in which the bending elasticity of chains is varied. The changes of the bending elasticity lead to various characteristics of the clustering behavior in the short-chain-only system. When a long self-propelled chain is mixed with these short chains, it can fold into the spiral-coil, a steadily rotating spiral conformation, either at a high density of short chains or at a low density if the long chain itself is sufficiently flexible. Our results qualitatively support the speculation on that the formation of the spiral-coil in the swarm of \emph{Vibrio alginolyticus} is triggered by collisions from the clusters of the shorter bacteria.
\end{abstract}

\maketitle
\newpage

\section{Introduction}
Collective locomotion such as bird flocks, fish schools and insect swarms is a common phenomenon in Nature. These beings autonomously travel together in a synchronous manner as if they were coordinated by intelligent communication \cite{Vicsek2012}. In simpler unicellular organisms, however, such collective behavior is also observed \cite{Zhang2010}. They are believed to rely on various cues and complex regulatory networks to achieve the coordination \cite{Dombrowski2004}. In the past few decades, this phenomenon has drawn a lot of attention from computational biologists, physicists, microbiologists and engineers for attempting to understand its origins and for the potential applications to technology \cite{Floreano2015}.

Among these researchers, Vicsek \emph{et al.} are usually considered as the pioneers to explore the physical properties of such systems \cite{Vicsek1995}. Additional to the agent-based rules that replicate the flocking behavior, the novelty of their model lies in the inclusion of a noise term, representing errors of agents' following the neighboring agents. This noise term and the density of the agents are crucial to the ferromagnetism-like phase transition. In the later research, self-propelled rods are extensively investigated for that they naturally achieve the alignment as the Vicsek model by the excluded volume interaction \cite{Peruani2006,Baskaran2008,Ginelli2010,Marchetti2013}. Self-propelled particles/rods are known to feature (i) giant fluctuations of local density \cite{Ramaswamy2003} and (ii) scale-free distribution of cluster sizes \cite{Heupe2004,Chate2008,Yang2010}. These systems are often compared with the experiments of bacterial locomotion as the minimal models because of the shape of the bacteria and the observed collective locomotion.

However, there is a great deal of diversity in terms of morphology and motility among bacterial species. As a result, rich dynamics different from self-propelled rods can be found in many living systems. The swarm of \emph{Vibrio alginolyticus} is one of such examples. While colonized on rich semi-solid surfaces, \emph{V. alginolyticus} cells elongate and express lateral flagella (transforming from monotrichous to peritrichous flagellation) for rapid surface motility \cite{McCarter2004,Kojima2007}. Therefore, at a high density these bacteria show active and dynamical patterns such as turbulence-like patterns \citep{Dombrowski2004,Wensink2012}. More importantly, the bacteria behave as having finite bending modulus rather than like rigid rods.

The mechanical properties of cell walls of certain bacterial species have been investigated in the literature. For instance, the longitudinal Young's modulus of the cell envelope of \emph{Escherichia coli} is determined as 50--150 MPa \cite{Tuson2012}, corresponding to a persistence length in the order of 10--100 m if the bacterium were an isotropic rod in thermal equilibrium at room temperature \cite{Broedersz2014}. Since \emph{V. alginolyticus} is also a Gram-negative germ like \emph{E. coli}, one would expect that its cell envelope should have a similar Young's modulus despite that no such measurement upon \emph{V. alginolyticus} has been reported. Nevertheless, bending with a radius of curvature in the order of 10--100 $\upmu$m has been observed in the swarm of \emph{V. alginolyticus}, especially during a collision between bacteria \cite{Lin2014}. The severe bending implies that some energy in the swarm must be transfered through collisions and temporarily stored in the form of elastic bending energy. Therefore, the bending elasticity of the cell envelope shall play a role in the in-plane rotational dynamics and the formation of clusters among the bacteria. Meanwhile, some of the very long bacteria ($> 100$ $\upmu$m) in the swarm fold into an ordered, compact conformation called the spiral-coil \cite{Lin2014} in spite of the stiff cell envelope. It is speculated that the onset of forming the spiral-coil is triggered by collision events from a number of clusters formed by the shorter bacteria.

Consequently, the bending elasticity and the number density of the bacteria would be the key factors that could determine the properties of clusters and in turn to influence the process of forming the spiral-coil in the swarm. In order to systematically investigate their effects upon the dynamics, it is most convenient to employ numerical simulation of self-propelled chains to represent the bacteria, and these two factors can be treated as parameters and easily tuned in the simulation.

Recently self-propelled semiflexible filaments have been numerically investigated \cite{Prathyusha2018,Duman2018}. These studies focus on monodisperse system and examine how the active transportation and the bending elasticity result in various phases of patterns, including a phase of rotating spiral-coils. The long-tailed distribution in length of the swarming \emph{V. alginolyticus} \cite{Lin2014}, however, may lead to different dynamical behavior from the monodisperse counterpart. For example, as we shall show below, the cluster sizes formed in the short-chain-only system is scale free. With the presence of a single long chain in the mixture system, on the other hand, this long chain introduces a specific length scale in the system and it changes the collective behavior of the short chains. In order to understand the formation of the spiral-coil in the bacterial swarm, we investigate a bidisperse system of self-propelled chains, of which one is long and the others are short, as the simplest representative example of the real bacterial swarm.

In this article we report the findings from the Brownian dynamics simulation of this bidisperse system. In the simulation each self-propelled chain is represented by linked beads moving through a viscous medium in a quasi-two-dimensional space, and the excluded volume interaction is the sole interaction between the chains. Specifically, we examine the generic properties of a group of short chains, and the spiral-coil formation in a mixture of a single long chain with the background short chains, whose bending modulus is deliberately fixed. The clusters formed by these chains are identified by the algorithm, density-based spatial clustering of applications with noise (DBSCAN)~\cite{Ester1996}.

The simulation results demonstrate that the spiral-coil is indeed spontaneously formed among the polar, self-propelled chains due to the combination of the steric interactions between the chains and their collective locomotion. The occurrence of such a conformation depends on the bending modulus of the long chain and the number density of the background short chains. At a low density the flexible long chain can easily fold into the spiral-coil due to its serpentine locomotion, but the semiflexible one only slightly deviates from the rod-like shape. On the other hand, at a high density there is no obvious difference in forming the spiral-coil between the flexible and semiflexible long chains. These findings qualitatively support the speculation on the formation of the spiral-coil and illustrate that complex dynamics in living systems can emerge from simple physical interactions.

The presentation of this research will be organized as follows. Section~\ref{sec:methods} introduces the model and methods used in this study. Section~\ref{sec:results} shows the results of the Brownian dynamics simulation, including the properties of the short-chain-only system and phase diagrams of the spiral-coil formation in the mixture system. Section~\ref{sec:discussion} discusses the theoretical analysis of the simulation results, and the mechanisms involved in the spiral-coil formation. Finally in Section~\ref{sec:conclusions} we conclude this article by summarizing what we have learned from this study regarding the formation of the spiral-coil and the generic properties of active matter.

\section{\label{sec:methods}Model and Methods}
Before we present the detail of the model and methods, we shall address the assumptions made in the present work. First we assume that hydrodynamic flows are irrelevant in this system. The locomotion of swarming bacteria are usually confined between a stress-free air-liquid interface and a porous liquid-solid interface (e.g. an agar plate). The thickness of the liquid layer is in the order of 1--10 $\upmu$m, comparable to the typical cell length of bacteria. These boundary conditions are drastically different from that of free-swimming bacteria. The flagella might interact with these boundaries, and the formation of the flagella bundle would be interrupted such that the force-dipole approximation of a swimming bacterium could break down; in such a scenario, a realistic model might involve a higher-order multipole or a multipole expansion, where the flow velocity should decay with the distance $r$ faster than that of the force-dipole, $v_{\text{f}}(r)\propto r^{-2}$. Experimental evidence shows that the hydrodynamic flow field indeed decays rapidly around the peritrichous bacteria in the proximity of a no-slip surface \cite{Cisneros2008,Drescher2011}. This suggests that persistent swimming and steric forces are predominant factors to determine the bacterial locomotion in the swarm.

Based on this assumption, the viscous medium merely dissipates energies through the drag, and the momentum is not conserved in the system. Such a system corresponds to the active \emph{dry} system discussed in Ref.~\cite{Marchetti2013}. In the meantime, we consider a regime of low Reynolds numbers ($\textrm{Re}$) in which the characteristic length scale is in the order of 1--100 $\upmu$m, the volumetric mass density and viscosity is the same as water, and the velocity is in the order of $10$ $\upmu$m/s, namely, $\textrm{Re}\sim 10^{-4}$ to $10^{-6}$.

Second, the absence of the hydrodynamic interaction means that the motility of the bacteria cannot be described by either the momentum transmission from the rotating flagella or the force-dipole approximation mentioned above. The simplest way to model the motile bacteria, therefore, is to assume the active propulsion acting on the cell body as if it were an \emph{external force} in the same way as the gliding bacteria \cite{Marchetti2013}. We further assume the propulsion uniformly distributes on the cell body, has a constant magnitude, and points along the local tangential direction towards one specific end as the leading end, which means that the chain is polarized. In the swarm of \emph{V. alginolyticus} there should be no physiological distinction for both ends of each bacterium and thus no preference to which way the bacterium is moving. However, the bacteria consistently move with one end as the leading end in the experiments except that occasionally a few bacteria fully stop and reverse the direction after meeting obstacles. Hence we reckon that it is appropriate to model the bacteria as polar chains to capture the motile behavior observed in the swarm.

Finally, we assume that between the chains there is \emph{only} the excluded volume interaction, which plays important roles in the dynamics of active matter \cite{Peruani2006,Baskaran2008,Ginelli2010,Marchetti2013}. We cannot entirely rule out other effects that resemble attraction in a real biological system, for example, the entanglement of lateral flagella in the swarm of \emph{V. alginolyticus}. Nevertheless, we specify that the exclude volume interaction is the only interaction between the chains in this work as it appears to be the most significant effect observed in the experiments \cite{Lin2014,Drescher2011}.

\subsection{Brownian Dynamics Simulation}
\begin{figure}
	\centering
    \includegraphics[width=\linewidth]{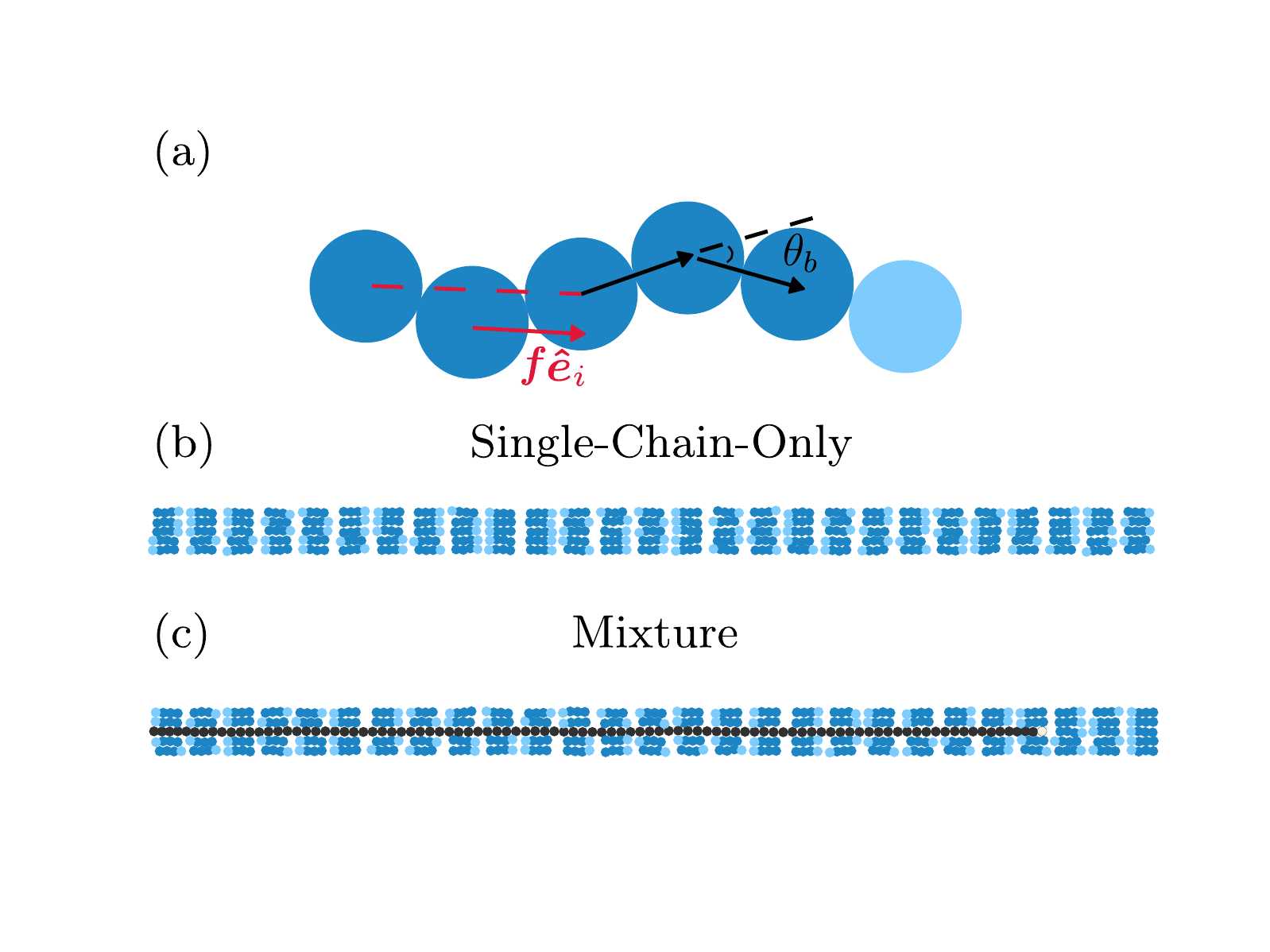}
    \caption{Schematic diagram of the model chain and snapshots of initial conditions. (a) The light blue (light gray) color notifies that the bead is the leading end. $\theta_{\text{b}}$ is the bending angle between two bonds, and $f$ is the magnitude of propulsion, acting along the local tangential direction $\bm{\hat{e}}_i$ towards the leading end. (b) and (c) are the zoom-in snapshots of initial condition for the single-chain-only system and the mixture system, respectively. Only five rows with 108 beads in each are shown. The long chain is colored in black and the white dot denotes its leading end.}
	\label{fig:chain-model}
\end{figure}

In this work a chain is modeled by $N$ connected beads, each of which has a diameter $b$. We are concerned with the position of beads in each chain over time, denoted as $\bm{R}_i(t)$ for the position of the bead $i\in\{1,2,\cdots,M\}$ with $M$ being the total number of beads. Fig.~\ref{fig:chain-model}(a) shows the schematic diagram of the model chain.

As we are examining a regime of low Reynolds numbers, the time evolution of $\bm{R}_i(t)$ is described by the over-damped Langevin equation
\begin{eqnarray}
	\zeta\diff{\bm{R}_i}{t}&=&\bm{f}_i(t)+\bm{\xi}_i(t)-\nabla_i U_{\text{b}}-\sum_{\substack{j=1\\j\ne i}}^M \nabla_i U_{\text{W}}(r_{ij})\nonumber\\& &-\sum_{\substack{j'\in\mathbb{I}_p\\j'=i-1,i+1}} \nabla_iU_{\text{F}}(r_{ij'}),
    \label{eq:Langevin}
\end{eqnarray}
where $\zeta$ is the friction constant, $\bm{f}_i(t)$ is the active propulsion exerted on the bead, $\bm{\xi}_i(t)$ is the random force, $\mathbb{I}_p$ is a set of bead indices designated to the chain $p$ such that $i\in\mathbb{I}_p$, and $U_{\text{b}}$, $U_{\text{W}}$ and $U_{\text{F}}$ are the interaction potentials that depend on the bead positions. Note that $r_{ij}$ (or $r_{ij'}$) denotes the separation between the bead positions $\bm{R}_i$ and $\bm{R}_j$ (or $\bm{R}_{j'}$). Since the bead is moving in a quasi-two-dimensional space and the hydrodynamic flow is absent, the bead follows Stokes' law and thus $\zeta=3\uppi b\eta$, where $\eta$ is the viscosity of water.

The active propulsion is simply $\bm{f}_i(t)=f\bm{\hat{e}}_i(t)$, where $f$ is the constant magnitude and the tangential unit vector $\bm{\hat{e}}_i(t)$ is defined as
\begin{equation}
	\bm{\hat{e}}_i(t)\equiv\frac{\bm{R}_{i+1}-\bm{R}_{i-1}}{\norm{\bm{R}_{i+1}-\bm{R}_{i-1}}}
\end{equation}
for the non-end beads. As for both of the ends, we define the unit vector $\bm{\hat{e}}_i^{\text{T}}\equiv(\bm{R}_{i+1}-\bm{R}_i)/\norm{\bm{R}_{i+1}-\bm{R}_i}$ for the tailing end and $\bm{\hat{e}}_i^{\text{L}}\equiv(\bm{R}_i-\bm{R}_{i-1})/\norm{\bm{R}_i-\bm{R}_{i-1}}$ for the leading end, respectively.

The random force $\bm{\xi}_i(t)$ is attributed to the rapid, random collisions of medium molecules upon the bead. It is characterized by a Gaussian probability distribution, as usual, with the zero mean $\langle\xi_{i\alpha}(t)\rangle=0$ and the correlation function $\langle\xi_{i\alpha}(t)\xi_{j\beta}(t')\rangle=2\zeta k_{\text{B}}T\delta_{ij}\delta_{\alpha\beta}\delta(t-t')$, where the index $\alpha$ and $\beta$ denote the component of the coordinates, $k_{\text{B}}$ is the Boltzmann constant, and $T$ is the temperature.

The interaction potentials represent three different kinds of interactions: the short-ranged excluded volume interaction ($U_{\text{W}}$), the bonding to maintain the integrity of the chain ($U_{\text{F}}$), and the bending elasticity ($U_{\text{b}}$). The excluded volume interaction is formulated by the Weeks-Chandler-Andersen (WCA) potential \cite{Weeks1971}
\begin{eqnarray}
 U_{\text{W}}(r) = \left\{ 
  \begin{array}{l l}
    4\varepsilon\left[\left(\frac{b}{r}\right)^{12}-\left(\frac{b}{r}\right)^{6}\right]+\varepsilon & \quad {\text{for}}\ r<2^{1/6}b\\
    0 & \quad {\text{for}}\ r\ge2^{1/6}b,
  \end{array} \right.
  \label{eq:WCA}
\end{eqnarray}
where $r$ is the separation between two beads and $\varepsilon$ is the well depth, which has been set to equal $k_{\text{B}}T$ in the simulation.

In this work we apply the finitely-extensible nonlinear elastic (FENE) model \cite{Kremer1990} instead of the simpler harmonic spring to the bond between beads in a chain. The FENE potential is 
\begin{eqnarray}
 U_{\text{F}}(r') = \left\{ 
  \begin{array}{l l}
    -\frac{1}{2}kR_0^2\log[1-(r'/R_0)^2] & \quad {\text{for}}\ r'<R_0\\
    \infty & \quad {\text{for}}\ r'\ge R_0,
  \end{array} \right.
\end{eqnarray}
where $r'$ is the separation between two connected beads, $k$ is the elastic constant, and $R_0$ defines the limit of the extensibility of bond. With a careful choice of $k$ and $R_0$, in addition to the WCA potential, the chain will robustly avoid crossing itself or the others.

Finally, the elastic bending energy is given by
\begin{equation}
	U_{\text{b}}=\frac{1}{2}k_{\text{l}}\theta_{\text{b}}^2,
\end{equation}
where the bending modulus $k_{\text{l}}=l_{\text{p}}k_{\text{B}}T/2b$ with $l_{\text{p}}$ being the persistence length \cite{Doi1988}, and $\theta_{\text{b}}$ is the bending angle. We calculate the bending angle between two consecutive bonds by $\theta_{\text{b}}=\cos^{-1}[\bm{R}'\cdot\bm{R}''/(\norm{\bm{R}'}\norm{\bm{R}''})]$, where $\bm{R}'=\bm{R}_i-\bm{R}_{i-1}$ and $\bm{R}''=\bm{R}_{i+1}-\bm{R}_i$, respectively.

\subsection{Simulation Conditions}
The simulation was employed with periodic boundary conditions in a square space of the size $L=240$ $\upmu$m. We arranged all the chains compactly into rows of beads as the initial condition, and each row contained 192 beads. The initial distance between beads was $1.12b$, the effective range of the WCA potential. The diameter of beads had been set to be $b=1$ $\upmu$m, but consecutive in-chain beads were allowed to overlapped during the simulation due to the combination of the WCA and FENE potential. All the chains were aligned parallel to the horizontal axis. The initial polarity of the single long chain ($N=96$) was fixed while that of the short chains ($N=4$) were randomly assigned. The long chain occupied half of the row at the center of the space and it was sandwiched by an equal number of rows of short chains on both sides. The other half of the central row was filled in with short chains as well. For the short-chain-only simulation, the whole central row was filled in with only short chains; see Fig.~\ref{fig:chain-model}(b) and (c).

The number density of the short chains can be expressed by either the number of chains per unit area $\rho$ or the fraction of projected area occupied by the short chains over the total area $L^2$. We will use the area fraction $C$ to express the density of the short chains. These two quantities can be easily converted to one another by $C=\uppi b^2\rho$ for $N=4$. In the simulation we set the number of short chains by the number of rows in the initial condition, as shown in Fig.~\ref{fig:chain-model}(b) and (c). We used $C=0.0$, 2.2, 4.3, 8.0, 13.2 and 26.3\% for the mixture system while in the short-chain-only simulation we used $C=26.4\%$.

The other key parameter of this research is the bending modulus $k_{\text{l}}$, or equivalently the persistence length $l_{\text{p}}$ under a fixed temperature. Here we categorize a chain with $l_{\text{p}}/b\ge N/2$ as semiflexible and $l_{\text{p}}/b<N/2$ as flexible, respectively, according to the difference between the contour length and the Kuhn length $2l_{\text{p}}$ \cite{Doi1988}. For the short-chain-only simulation, we tested $l_{\text{p}}/b=2$, 20, 40, 80, 120 and 160. As the cluster properties could depend on the bending elasticity of the short chains, we would prefer to reduce the complication in the mixture system in order to focus on examining the dynamics of the long chain. As a result, for the mixture system, the persistence length of the short chains was kept as $l_{\text{p}}/b=20$ or 120 while that of the long chain was set as $l_{\text{p}}/b=2$, 20, 40, 60, 80, 100, 120, 160 and 200 to cover both the flexible and the semiflexible regime.

The thermal fluctuations were subject to the thermal energy $k_{\text{B}}T=4.12$ pN$\cdot$nm, corresponding to a temperature $25^\circ$C. The friction constant $\zeta$ was given by the Stokes' drag with the water viscosity $\eta=8.9\times 10^{-4}$ Pa$\cdot$s at $25^\circ$C. The propulsion magnitude $f$ was kept at 0.2 pN based on the experimental estimation of the flagellar motor thrusting force \cite{Chattopadhyay2006} and the density of flagella on a swarming \emph{V. alginolyticus}. The elastic constant $k=0.742$ pN/$\upmu$m and $R_0=1.25$ $\upmu$m were set for the FENE potential throughout the simulation.

\subsection{Cluster Identification}
DBSCAN is a clustering algorithm based on local density and allows the existence of noise points. The minimum local density is determined by two parameters: the searching radius and the minimum points inside the searching circle. The algorithm starts from an arbitrary point, then the searching is spread out to other neighboring points to reach the required density. After scanning all of the points, the connected searching circles are regarded as one cluster, and the points with a local density lower than the minimum density are the noises. The clustering is based on the expansion of small searching circles, thus the cluster can be any shape composed by searching circles.

The cluster identification in our system is a spatiotemporal problem, in which self-propelled chains gather together in space, and move towards the same direction. DBSCAN plays the role of spatial clustering method. However, with only spatial information we cannot distinguish two close clusters with the opposite direction of motion. As a result, in the process of expanding DBSCAN cluster, we also required that the angular deviation of the velocity between neighboring chains had to be less than 30${}^{\circ}$. In the cluster identification we used the center of mass to represent the position of each chain. Meanwhile, in order to identify the free-roaming chains, namely, individual chains not being parts of any cluster, we deliberately set the minimum points as 2. In this way, the noise point in the original DBSCAN would be regarded as a cluster composed of one single chain. For the searching radius, we set it as 2 $\upmu$m as a tolerant condition. Finally, the direction of velocity was averaged over a time window of 50 ms. 

\subsection{Criteria of Identifying Spiral-Coils}
We define the spiral-coil as a spiral conformation that steadily rotates on a time scale exceeding other characteristic times of the dynamics, such as the characteristic rotation time $\tau_{\text{r}}$ (see Section~\ref{sec:results}). The identification of a spiral-coil from other conformations is crucial to investigating the effects of $k_{\text{l}}$ and $C$ upon its formation. Although an ideal spiral-coil can be mathematically approximated by the Archimedean spiral, in practice the long chain does not always form a perfect one, especially with a high area fraction of the short chain. In order to proceed the investigation, we identify the spiral-coil by the criteria as follows.

First, its radius of gyration $R_{\text{g}}$ shall be less than 20 $\upmu$m and remain constant over time. In addition to the perfect configuration, a spiral-coil sometimes encloses a number of the short chains between the rounding threads. This will increase the size of the spiral-coil to be larger than an Archimedean spiral. Since a spiral must be smaller than the circle in the same contour length, we set this upper limit for $R_{\text{g}}$ based on the circular conformation of the long chain with a margin of tolerance.

Second, its end-to-end distance shall remain constant over time. In some cases the conformation is not a spiral yet it has a $R_{\text{g}}$ that passes the criterion above. Nonetheless, the end-to-end distance of these cases usually varies in different ways from that of the spiral-coil. The end-to-end distance of a spiral-coil measures the distance between the leading end trapped at the center and the tailing end. Since we expect that the conformation is maintained for a long time, its end-to-end distance shall be constant as well. This criterion excludes those compact conformations of which the leading end is not trapped inside the spiral.

Third, its end-to-end vector shall rotate at a constant angular velocity over time. The angular velocity is measured by plotting the cumulative polar angle of the end-to-end vector versus time as a continuous curve. We quantify the stability of the angular velocity by the Pearson correlation of the cumulative polar angle and the time. The correlation coefficient during the rotation of a spiral-coil shall be greater than 0.98.

Finally, a conformation shall pass all of the three criteria for a time span of one second or more to be identified as a spiral-coil.

\section{\label{sec:results}Results}
\subsection{Dynamics of Short-Chain-Only System}
\begin{figure}
\centering
\includegraphics[width=\linewidth]{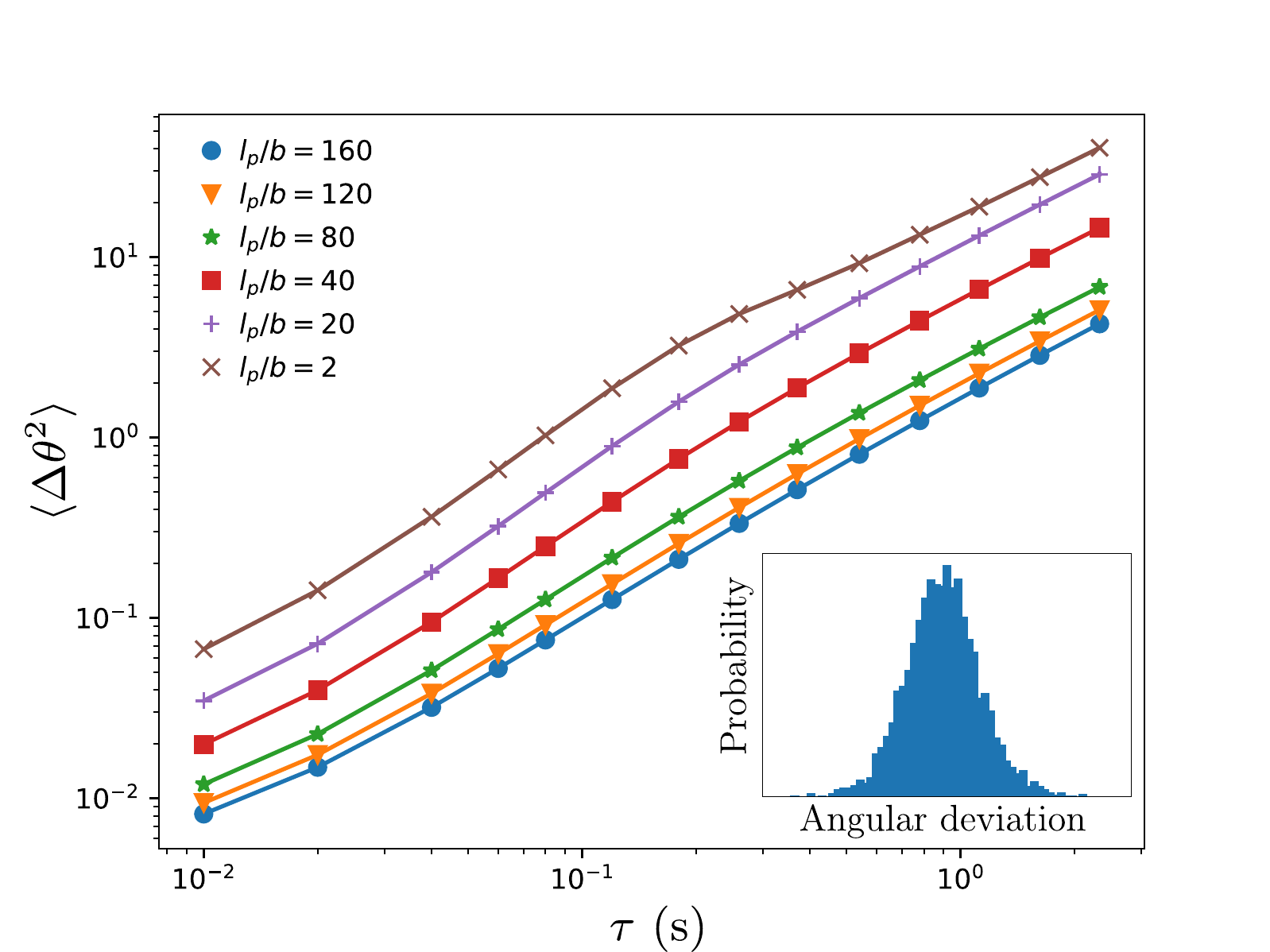}
\caption{Mean-squared angular deviation of the end-to-end vector of the short chains. $\langle\Delta\theta^2\rangle$ can be well fitted by the linear relationship $\langle\Delta\theta^2\rangle=2D_{\text{r}}\tau$, where $D_{\text{r}}$ decreases with the increasing $l_{\text{p}}$. The inset shows that the distribution of the angular deviation about the straight conformation resembles a Gaussian distribution.}
\label{fig:MSR}
\end{figure}
We first investigate the collective motility of the short-chain-only system. We aim to learn the cluster formation in these monodisperse system as the basis before we move on to examine the bidisperse mixture system. Even though we focus on specific conditions that are relevant to the bacterial swarm, some of the results presented in this subsection can be directly compared with those presented in Refs.~\cite{Prathyusha2018,Duman2018}. More detailed information regarding the dynamics of self-propelled semiflexible filaments is given in these two papers.

\begin{figure}
\centering
\includegraphics[width=\linewidth]{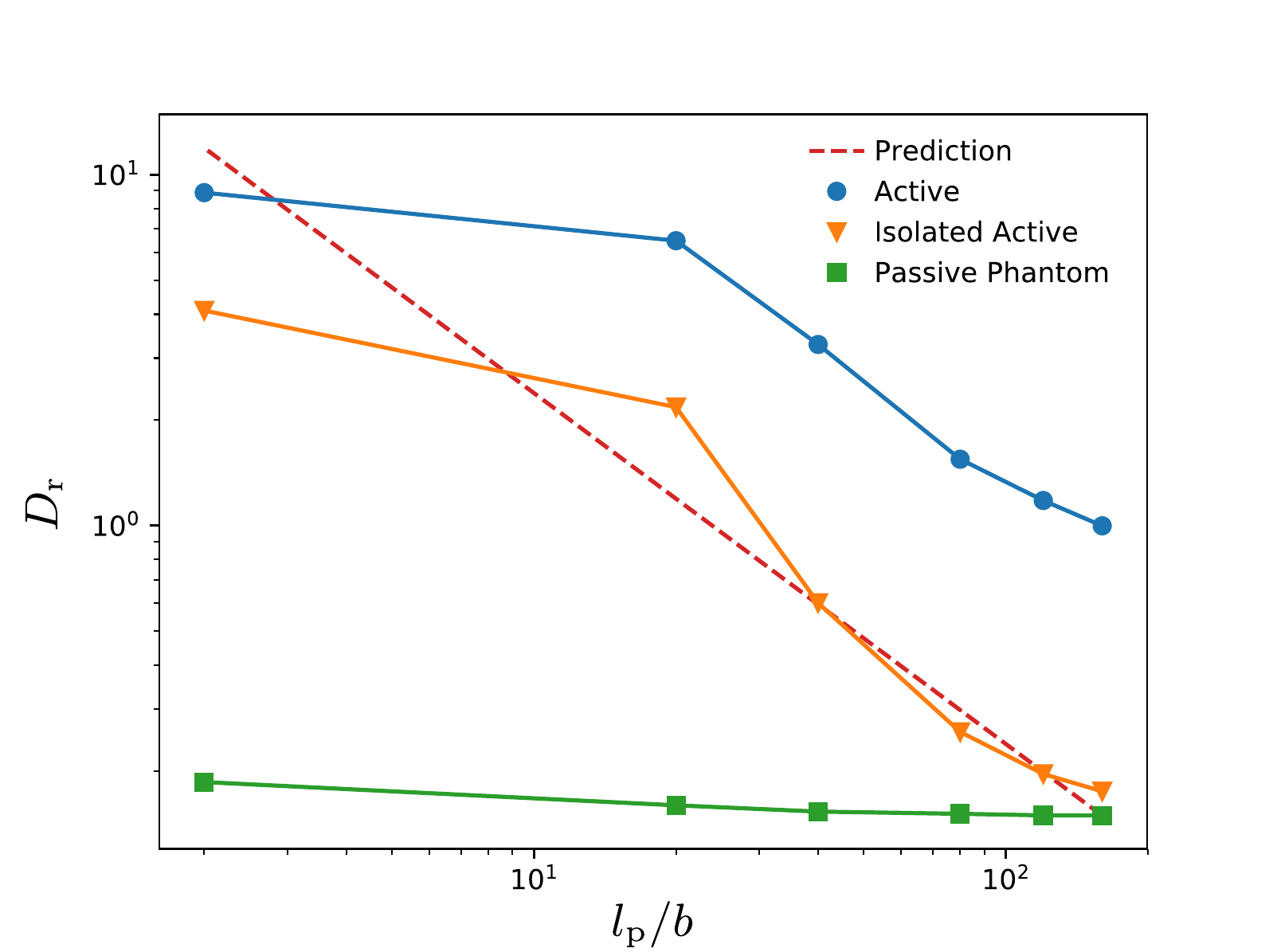}
\caption{Rotation rate of chains with the corresponding persistence length. $D_{\text{r}}$ is obtained by fitting $\langle\Delta\theta^2\rangle=2D_{\text{r}}\tau$ from three types of simulation conditions: \emph{Active} referring to the typical clustered short chains as shown in Fig.~\ref{fig:MSR}, \emph{Isolated Active} referring to isolated self-propelled short chains, and \emph{Passive Phantom} referring to isolated short chains without the propulsion. The prediction of Eq.~(\ref{eq:D_r-theory}) agrees with the isolated active chains very well except the very flexible ones.}
\label{fig:Dr-fit}
\end{figure}

Fig.~\ref{fig:MSR} is a log-log plot that shows the mean-squared angular deviation of the end-to-end vector $\langle\Delta\theta^2\rangle$ versus the lag time $\tau$. The data can be well fitted by $\langle\Delta\theta^2\rangle=2D_{\text{r}}\tau$, where $D_{\text{r}}$ is the rotation rate. The inset of Fig.~\ref{fig:MSR} shows that the orientation of the end-to-end vector approximately follows a Gaussian distribution, where the center corresponds to the straight conformation. It is notable that the rotation rate $D_{\text{r}}$ decreases with the increasing persistence length $l_{\text{p}}$, and it is one to two orders larger than that of passive, phantom chains (see Fig.~\ref{fig:Dr-fit}). These results suggest that the self-propelled chains significantly deviate from the prediction of the equipartition theorem \cite{Broedersz2014}.

\begin{figure}
\centering
\includegraphics[width=\linewidth]{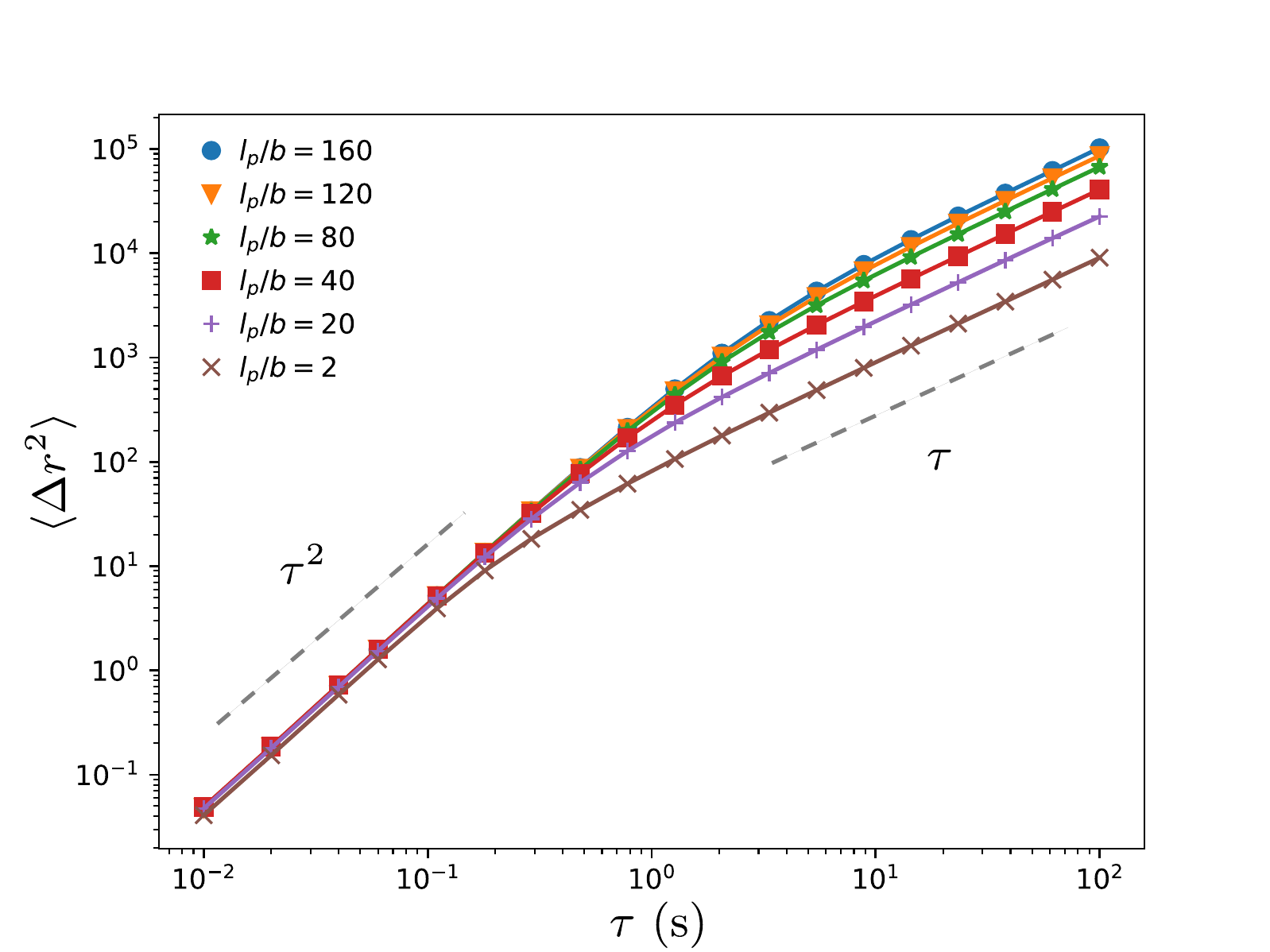}
\caption{Mean-squared displacement of the center of mass of the short chains. The MSD clearly crosses over from a ballistic regime ($\propto\tau^2$) to a diffusive one ($\propto\tau$). Note that in the diffusive regime the curves separate from each other with the various $l_{\text{p}}$. The MSD can be qualitatively described by Eq.~(\ref{eq:MSD-prsisBrownian}).}
\label{fig:MSD}
\end{figure}

The mean-squared displacement (MSD) of the center of mass $\langle\Delta r^2\rangle$, shown as a log-log plot in Fig.~\ref{fig:MSD}, exhibits a crossover from the ballistic movement on the short time scales to the diffusive motion on the long time scales. Interestingly, the behavior of these short chains is very similar to the persistent random walk of active Brownian particles in two dimensions. The MSD in Fig.~\ref{fig:MSD} qualitatively follows the relationship
\begin{eqnarray}
 \langle\Delta r^2\rangle = \left\{ 
  \begin{array}{l l}
    4D\tau+\langle \bm{u}^2\rangle \tau^2 & \quad {\text{for}}\ \tau\ll\tau_{\text{r}}\\
    \left(4D+\frac{\langle \bm{u}^2\rangle}{D_{\text{r}}}\right)\tau & \quad {\text{for}}\ \tau\gg\tau_{\text{r}},
  \end{array} \right.
  \label{eq:MSD-prsisBrownian}
\end{eqnarray}
where $\bm{u}$ is the velocity, $D$ is the translational diffusion coefficient, and $\tau_{\text{r}}\equiv 1/D_{\text{r}}$ is the characteristic rotation time \cite{Elgeti2015}. In our system the active propulsion dominates over the translational diffusivity (characterized by the P\'{e}clet number $\textrm{Pe}\sim 10^2$ for the short chain), hence the quadratic term is dominant in the regime $\tau\ll\tau_{\text{r}}$. This crossing-over behavior is consistent with the results shown in Refs.~\cite{Prathyusha2018,Duman2018}.

\begin{figure}
\centering
\includegraphics[width=\linewidth]{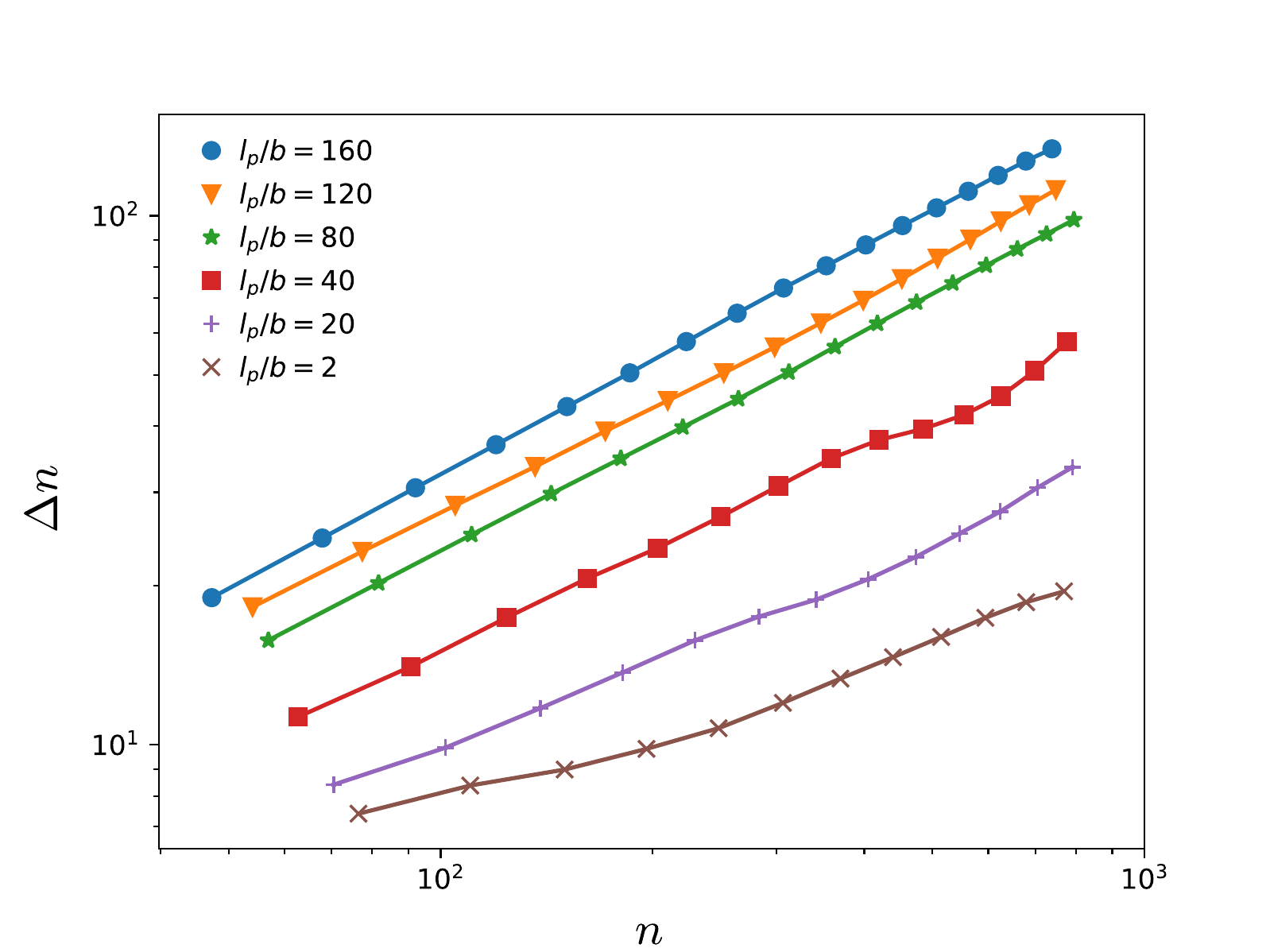}
\caption{Fluctuations of the local number of chains. At a high $l_{\text{p}}$ the short chains exhibit the giant number fluctuations as the polar self-propelled rods. However, softer chains do not appear to follow a simple power law as their stiffer counterparts.}
\label{fig:num_fluc}
\end{figure}

Like many other systems of self-propelled objects, the simulated system also features the giant number fluctuation of local chains under certain conditions. In thermodynamic equilibrium the variance of local chain numbers shall scale with the mean number over an arbitrary area of linear size $a$ as $\Delta n^2=\langle[n(a)-\bar{n}(a)]^2\rangle\sim\bar{n}(a)^{2\gamma}$, where $\bar{n}(a)=\rho a^2$ and $\gamma=1/2$. In an active system it is usually found that $\Delta n$ scales with $\bar{n}(a)$ faster than $1/2$ \cite{Ramaswamy2003}. Fig.~\ref{fig:num_fluc} demonstrates the scaling behavior of $\Delta n$ versus $\bar{n}(a)$ with various persistence length $l_{\text{p}}$ under the same area fraction. With the increasing $l_{\text{p}}$, the exponent $\gamma$ monotonically increases from $0.55\pm0.01$ ($l_{\text{p}}=20b$) to $0.7158\pm0.0007$ ($l_{\text{p}}=160b$). The result with the longest persistence length is close to the predicted exponent $0.8$ for polar self-propelled rods \cite{Chate2008,Peruani2012}, which are not bendable at all (i.e. $l_{\text{p}}\rightarrow\infty$). In the meantime, the fitted $\gamma$ is consistent with the result of the intermediate P{\'e}clet number shown in Fig. 10 of Ref.~\cite{Duman2018}. Another notable point shown in Fig.~\ref{fig:num_fluc} is the curve of the softer chains ($l_{\text{p}}=2b,20b$ and $40b$) do not appear to follow a simple power law; the attempt of fitting the curve for the case $l_{\text{p}}=2b$ resulted in an anomalous exponent $\gamma=0.30\pm0.01$.

\begin{figure}
\centering
\includegraphics[width=\linewidth]{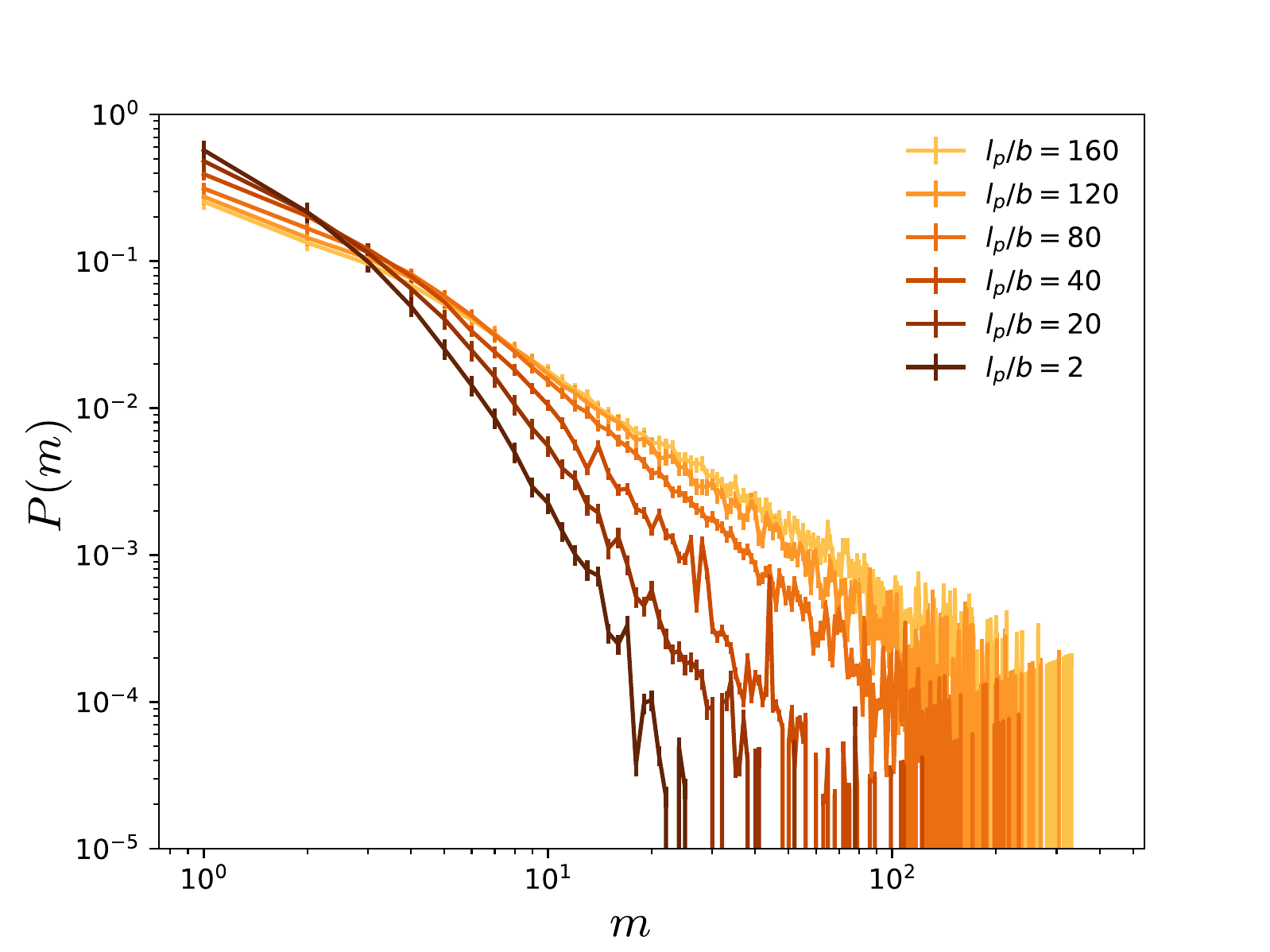}
\caption{Cluster size distribution of the short chains. The distribution does not follow a simple power-law decay as the self-propelled particles/rods reported in the literature. Nonetheless, for the intermediate to large clusters the distribution can be fitted by power-laws with a varying exponent depending on $l_{\text{p}}$. The softer the chains are, the less likely one finds them forming large clusters.}
\label{fig:CSD}
\end{figure}

Another important feature found in the simulated system is the scale-free distribution of cluster sizes. The clusters were identified by the algorithm DBSCAN, as described in Section~\ref{sec:methods}. Fig.~\ref{fig:CSD} shows a log-log plot of the probability of finding a cluster that contains $m$ chains, $P(m)$. Probability distribution of cluster sizes following a simple power law has been observed in the simulation of self-propelled particles/rods \cite{Heupe2004,Chate2008,Yang2010}. On the other hand, Peruani \emph{et al.} study collective motion in colonies of gliding bacteria \emph{Myxococcus xanthus} and propose that the steady-state cluster size distribution, equivalent to $P(m)$ here, can be fitted by the form $P(m)\propto m^{-\gamma_0}\exp(-m/m_0)$, where $m_0$ is supposed to be a function of area fraction \cite{Peruani2012}. Neither of the proposed form fits our results as each curve in Fig.~\ref{fig:CSD} is characterized by two distinct exponents for small and large clusters, respectively. Determining the exact scaling behavior would require thorough investigation over a large extent of the parameter space, which is beyond the scope of this article. What being relevant in Fig.~\ref{fig:CSD} is that the decay of $P(m)$ depends on the persistence length $l_{\text{p}}$; fitting for the intermediate and large cluster sizes ($m>3$) by $P(m)\propto m^{-\gamma_{\text{c}}}$ results in $\gamma_{\text{c}}=1.30\pm0.04$ ($l_{\text{p}}=160b$) and $\gamma_{\text{c}}=2.70\pm0.10$ ($l_{\text{p}}=2b$), respectively. Therefore, with the decreasing persistence length, it is less likely to find the chains forming large clusters. 

\subsection{Mixture of Single Long Chain and Background Short Chains}

\begin{figure}
	\centering
	\includegraphics[width=\linewidth]{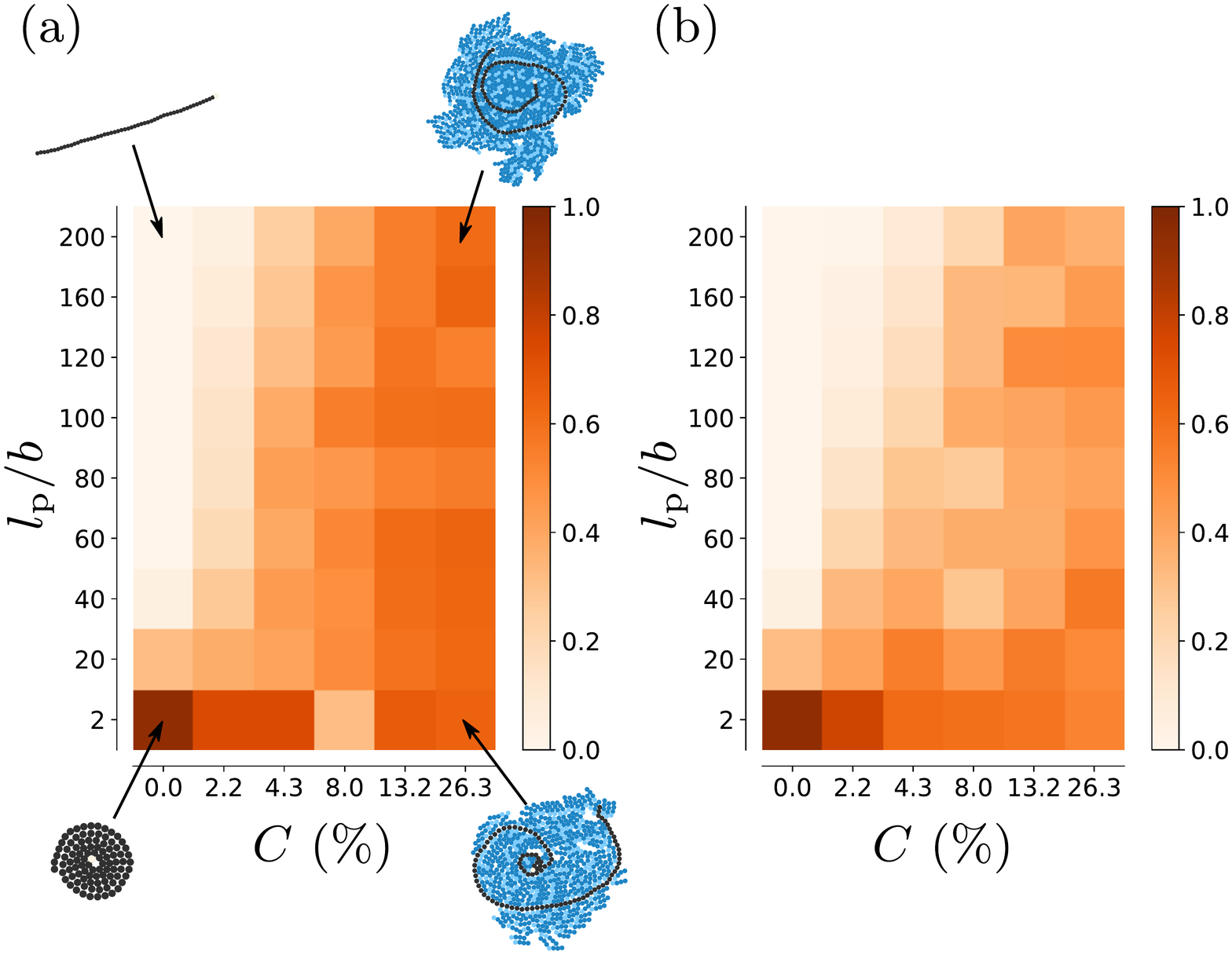}
	\caption{Phase diagrams of the probability of forming the spiral-coil and snapshot of a few representative long chains. The color of each block notifies the probability corresponding to the persistence length $l_{\text{p}}$ of the long chain and the area fraction $C$ of the background short chains. Note that the persistence length of the background short chains is fixed as (a) $l_{\text{p}}=120b$ and (b) $l_{\text{p}}=20b$, respectively, throughout the simulation of the mixture.}
	\label{fig:phase_diagram}
\end{figure}

Fig.~\ref{fig:phase_diagram} demonstrates the phase diagrams of the probability of forming the spiral-coil and the snapshot of a few representative conformations of the long chain for different conditions. The left two snapshots in Fig.~\ref{fig:phase_diagram}(a) are of a single long chain without any background short chain. The flexible long chain is able to self-fold into the spiral-coil \cite{Isele-Holder2015} due to its very low bending modulus and the serpentine locomotion. However, with the increasing bending modulus, it is more difficult for the long chain to fold itself. The right two snapshots in Fig.~\ref{fig:phase_diagram}(a) represent a common scenario at large $C$ in which the long chain is surrounded by a large number of the short chains, transversely and rotationally moving together as a cluster. In comparison with the spiral-coil in the monodisperse system, the topological constraints due to the surrounding short chains is not as drastic as the surrounding long chains in Ref.~\cite{Duman2018}, because individual short chains are able to move away from the clusters by collisions or the rotational diffusivity.

It is notable that even though the persistence lengths in the right two cases are significantly different, the semiflexible chain exhibits a more compact conformation than the flexible one. In particular, the local radius of curvature of the semiflexible chain is an order smaller than its persistence length $l_{\text{p}}=200b$. As the thermal fluctuation in the lateral direction (perpendicular to the tangential vector of the curvilinear contour) is too weak compared with the required bending energy, and there is no attraction between two arbitrary beads, this result suggests that the substantial bending is attributed to the collective motion of the surrounding short chains.

Following the criteria of identifying the spiral-coil in Section~\ref{sec:methods}, we analyze the simulation data and produce the phase diagrams featuring the probability of forming the spiral-coil under various $l_{\text{p}}$ and $C$. This probability is obtained by computing the proportion of chains that pass the criteria of the spiral-coil in the ensemble, of which the size for each simulation condition is 120. Fig.~\ref{fig:phase_diagram} shows the phase diagrams as color maps, where in (a) $l_{\text{p}}/b=120$ and in (b) $l_{\text{p}}/b=20$ for the background short chains. The color illustrates the probability of forming the spiral-coil with the corresponding parameters. Semiflexible long chains are able to fold into the spiral-coil with a sufficiently large number of background short chains ($C\ge 8.0\%$), and the probability generally increases with the increasing number of the short chains. On the other hand, flexible long chains with $l_{\text{p}}\le 20b$ can form the spiral-coil with very few number or none of the short chains. With a large number of background short chains ($C=13.2$ and 26.3\%), the probability of folding into a spiral-coil is similar between the flexible and the semiflexible long chains. Overall the trend in both Fig.~\ref{fig:phase_diagram}(a) and (b) is qualitatively the same, but in the case of Fig.~\ref{fig:phase_diagram}(b) the boundary of spiral-coil formation shifts to higher $C$ and the probabilities of forming the spiral-coil are lower than that of Fig.~\ref{fig:phase_diagram}(a).

More interestingly, for the chain with $l_{\text{p}}=2b$ the probability of folding into the spiral-coil first decreases and then increases with the increasing number of background chains in Fig.~\ref{fig:phase_diagram}(a); the minimal probability occurs at $C=8.0\%$. This might be explained by the fact that the surrounding short chains mediate the excluded volume interaction as if it were a long-ranged repulsion for the long chain. This process is analogous to the swollen polymer chain in a good solvent above the $\Theta$-temperature \cite{Doi1988}; the leading end of the swollen long chain is less likely to fold into its own contour, resulting in a lower probability of forming the spiral-coil. However, this argument requires further investigation to explain why the dip is absent in the panel (b), and why the stiffer long chains do not seem to be affected in the same manner.

\section{\label{sec:discussion}Discussion}
In Section~\ref{sec:results} we have learned that the cluster formation depends on the persistence length $l_{\text{p}}$. We also have learned that $l_{\text{p}}$ plays an important role in determining the rotational behavior in the short-chain-only system. In this section we attempt to analyze the process of forming clusters by considering the balance of chain fluxes in and out of a cluster, followed by the discussion on why the area fraction and possibly the bending modulus of the background short chains are such important factors to the formation of the spiral-coil.

\subsection{Formation of clusters}
Let us consider an arbitrary cluster of $m$ self-propelled chains that is far away from other clusters. Each of these chains is constituted of $N$ linked beads. Therefore, the contour length of the self-propelled chain is $l=N b$. For simplicity let us assume that the cluster is in a circular shape with a radius $R_{\text{c}}$. The area $A$ of the cluster is given by $A=\uppi R_{\text{c}}^2=mN\uppi b^2$, thus $R_{\text{c}}=b\sqrt{N m}$. The time evolution of the chain number in the cluster can be calculated by considering the influx and the efflux of chains.

The influx is provided by the free-roaming chains around the cluster. Assuming that in a region $\Omega$ the \emph{free-roaming} chains are uniformly distributed, with an area density $\rho_{\text{f}}(t)$, between the cluster periphery and the boundary of $\Omega$. The rate of joining in the cluster for each free-roaming chain in this region is $lu/b^2$, where $u$ is the speed of the free-roaming chain. Although the orientation of the free-roaming chains relative to the movement of the cluster would affect the probability of merging with the cluster, on average this should only contribute to a pre-factor as the orientation of the free-roaming chains is expected to be uniformly random. To avoid complication let us consider the region $\Omega$ with a radius $R_{\text{c}}+l$ so that most of the free-roaming chains are guaranteed to merge with the cluster. Therefore, the influx is given by
\begin{equation}
	J_{\text{I}}\sim \frac{\uppi lu}{b^2}[(R_{\text{c}}+l)^2-R_{\text{c}}^2]\rho_{\text{f}}(t).\label{eq:influx}
\end{equation}

The efflux is provided only by the peripheral chains in the cluster, because the inner chains are trapped by the excluded volume interaction. More precisely, due to the uniaxial geometry of chains and the resulting alignment, only the peripheral side chains contribute to the efflux. The number of these peripheral side chains can be estimated as $R_{\text{c}}/l$, and the rate of departing from the cluster is simply their mean angular velocity $\bar{\omega}$. Hence efflux is given by
\begin{equation}
	J_{\text{E}}\sim \bar{\omega}\frac{R_{\text{c}}}{l}.\label{eq:efflux}
\end{equation}
If the self-propelled chain were a rigid rod, then the mean angular velocity would be simply the angular drift velocity corresponding to the rotational diffusion of the rod. As for the flexible and the semiflexible chains considered here, $\bar{\omega}$ should depends on the temperature $T$ and the bending modulus $k_{\text{l}}$, and it can be calculated from the root-mean-squared angular deviation obtained in the simulation.

Combining Eqs.~(\ref{eq:influx}) and (\ref{eq:efflux}), the time evolution of the number of chains in the cluster is written as
\begin{equation}
	\diff{m}{t}=G\left[2\sqrt{N}\rho_{\text{f}}(t)-\frac{\bar{\omega}}{G\sqrt{N}}\right]\sqrt{m}+GN\rho_{\text{f}}(t),
\end{equation}
where $G=\uppi N^2bu$.

For a steady-state cluster size $m^*$, the density threshold $\rho_{\text{f}}^*$ can be derived by considering the balance between the influx and the efflux as $J_{\text{I}}=J_{\text{E}}$. Hence,
\begin{equation}
	\rho_{\text{f}}^*=\frac{\bar{\omega}}{(2\sqrt{m^*}+\sqrt{N})G}\left(\frac{\sqrt{m^*}}{N}\right).
\end{equation}
Whenever $\rho_{\text{f}}(t)<\rho_{\text{f}}^*$, the cluster will begin to reduce its size until a new steady-state is reached. This result suggests that a cluster of flexible chains will require a higher $\rho_{\text{f}}$ around it to sustain its size, because flexible chains have a higher mean angular velocity $\bar{\omega}$. The mean angular velocity $\bar{\omega}$ can be estimated in the following way: Fig.~\ref{fig:MSR} shows the mean-squared angular deviation $\langle\Delta\theta^2\rangle=2D_{\text{r}}\tau$. Therefore the root-mean-squared angular deviation is given by $\Delta\theta_{\text{rms}}=\sqrt{2D_{\text{r}}\tau}$. By inserting the shortest time scale in the simulation, namely, the size of a time step $h$, into the lag time $\tau$ we obtain the mean angular velocity as
\begin{equation}
	\bar{\omega}=\frac{\Delta\theta_{\text{rms}}}{h}=\sqrt{\frac{2D_{\text{r}}}{h}}.
\end{equation}
This is consistent with the results of our simulation shown in Fig.~\ref{fig:CSD} as those flexible chains are less likely forming large clusters than the stiffer ones under the same condition. In turn these smaller clusters reduce the capability of short chains to fold the long chain, as demonstrated in Fig.~\ref{fig:phase_diagram}(a) and (b).

However, analytically deriving the general form of $\rho_{\text{f}}(t)$ is difficult for that chain exchanges also happen across the boundary of $\Omega$. In fact over the whole space the area density of free-roaming chains should have been written as a function of position and time as $\rho_{\text{f}}(\bm{x},t)$. In an active system it is known that the number fluctuation can be significant as demonstrated in Fig.~\ref{fig:num_fluc}, thus the area density of free-roaming chains will not be uniform for a wide range of length scales. Moreover, the spatial and temporal evolution of $\rho_{\text{f}}(\bm{x},t)$ also depends on the size and the position of clusters, and the interactions between clusters such as merging two clusters after collision further complicates the calculation. A detailed derivation of $\rho_{\text{f}}(\bm{x},t)$ is far beyond the scope of this article. Nonetheless, based on the results presented in Section~\ref{sec:results} and the analysis above, we can summarize that the persistence length and the density of the self-propelled chains indeed play major roles in determining the size of clusters.

\subsection{\label{sec:sub-bending}Bending of the Long Chain}
A perfect spiral-coil can be approximately represented by the Archimedean spiral in the polar coordinates (except near the origin due to the volume exclusion) as $r_{\text{A}}=b\phi/2\uppi$, where $r_{\text{A}}$ is the radial distance and $\phi$ the polar angle. According to the worm-like chain model, the elastic bending energy per unit arc length of the Archimedean spiral is
\begin{equation}
	\delta E_{\text{b}}^{\text{u}}(\phi)=\frac{\uppi^2l_{\text{p}}k_{\text{B}}T}{b^2}\left[\frac{(\phi^2+2)^2}{(\phi^2+1)^3}\right],
    \label{eq:e-bending-u}
\end{equation}
which is monotonically decreasing with the increasing $\phi$, and it can be seen as the upper bound of the elastic bending energy per unit arc length.

Meanwhile, the lower bound can be taken by considering the long chain to adopt a circular conformation $\Pi$, because folding into a spiral-coil \emph{always} requires smaller radii of curvature than the radius of such a circle. Hence the lower bound of the elastic bending energy per unit arc length to form a spiral-coil is given by
\begin{equation}
	\delta E_{\text{b}}^{\text{l}}=\frac{\uppi^2l_{\text{p}}k_{\text{B}}T}{N^2 b^2}.
\end{equation}

For isolated long chains, $\delta E_{\text{b}}^{\text{l}}$ is a particularly informative measure to show why the very soft chain is able to self-fold into a spiral-coil. In these cases (i.e. $C=0.0\%$) the thermal fluctuation perpendicular to the local tangential direction contributes to the bending, whilst the active propulsion is acting along the local tangential direction and has no contribution at all. The total bending energy of the circular chain is $E_{\text{b}}^{\text{l}}=\delta E_{\text{b}}^{\text{l}}N b$. Comparing this quantity with the thermal energy scale $\frac{1}{2}k_{\text{B}}T$, one shall find that for a chain to bend into a circle solely by the thermal fluctuation, the persistence length has to be smaller than $N b/2\uppi^2$. Taking the long chain in the simulation ($N=96$) as an example, the persistence length of a self-folding spiral-coil would be $l_{\text{p}}<4.86b$.

In the other cases with the presence of the short chains, the bending of stiffer chains must be facilitated by collisions between the long chain and clusters of short chains. Whenever the long chain is undergoing the spiral-coil formation, the local bending energy $\delta E_{\text{b}}(s,t)$ must be between the interval $\delta E_{\text{b}}^{\text{l}}<\delta E_{\text{b}}(s,t)<\max\{\delta E_{\text{b}}^{\text{u}}(\phi)\}$, where $s$ is the curvilinear coordinate along the long chain. It is notable that $\delta E_{\text{b}}^{\text{l}}$ is $\sim N^2$ times smaller than $\max\{\delta E_{\text{b}}^{\text{u}}(\phi)\}$; the longer a chain is, the easier the chain folds into the spiral-coil. The local bending energy of a formed spiral-coil, however, primarily depends on the persistence length, because the term within the brackets of Eq.~(\ref{eq:e-bending-u}) is up to the order of unity. Furthermore, the long chain sometimes can be trapped by its own conformation, and the excluded volume interaction between the segments sustain the energetically costly bending.

\subsection{Serpentine Locomotion}
The most distinguishable characteristics of a self-propelled chain from the self-propelled particles or rods is the serpentine locomotion, which displays seemingly random curved trajectories of displacement. Such a locomotion is resulted from the combination of the bending flexibility, the tangential propulsion along the conformation, and the lateral thermal fluctuation of the leading end. This is analogous to the reptation (a term derived from the word \emph{reptile}) of polymer melts, where each polymer chain diffusively moves to and fro along a curvilinear tube due to the topological constraints of neighboring polymers \cite{Doi1988}. The self-propelled chain, however, exhibits the curvilinear conformation and locomotion because each segment actively follows the next segment in the front up to the leading end, which is randomly exploring a circular sector in its front. In other words, this is truly a snake-like motion.

The  major consequence of the serpentine locomotion upon the dynamics of self-propelled chains is as follows. The leading end freely explores a circular sector in its front unless there is obstacle blocking the way. Hence these chains change their courses even without colliding into other chains. How much the leading segment deviates from the straight conformation is determined by the lateral thermal fluctuation against the bending elasticity. Without long-ranged interaction or force from the other chains, the correlation of the unit tangential vector $\bm{q}(s,t)$ between the curvilinear position $s$ and $s'$ shall be $\langle\bm{q}(s,t)\cdot\bm{q}(s',t)\rangle=\exp(-|s-s'|/l_{\text{p}})$. Since the chain is in a regime where the active transportation is dominant against the translational diffusion, one can estimate the rotation rate as
\begin{equation}
	D_{\text{r}}=\frac{1}{\tau_{\text{r}}}=\frac{u}{l_{\text{p}}}\approx\frac{f}{3\uppi\eta bl_{\text{p}}}.
    \label{eq:D_r-theory}
\end{equation}
This prediction agrees with the simulation of isolated short chains very well for high persistence lengths, as shown in Fig.~\ref{fig:Dr-fit}. However, it overestimates the value for the flexible chains. The discrepancy might be caused by the volume exclusion between beads, which reduces the bending as if it were increasing the persistence length.

In the cases where $l_{\text{p}}\rightarrow\infty$ or $f\rightarrow 0$, $\tau_{\text{r}}$ becomes too long and irrelevant, and the rotational diffusivity would instead determine the rotational relaxation time. For example, the rotational relaxation time for an isolated self-propelled rod (assuming propulsion acting along the long axis) \cite{Doi1988} on a plane is expected to be
\begin{equation}
	\tau_{\text{r}}'\sim\frac{\eta l^3}{k_{\text{B}}T\log(l/b)},
\end{equation}
which would be significantly longer with a long contour length than that of the self-propelled chain.

\subsection{Trapped Leading End and the Resulting Spiral-Coil}
The serpentine locomotion can be drastically influenced by the presence of other self-propelled chains. For example, when the leading end collides into an obstacle (e.g. a cluster of chains, or even the chain itself), the propulsion from the segments behind temporarily bend the bonds, and the leading end turns to move along the interface of the obstacle until it deviates away from the interface on the time scale $\tau_{\text{r}}$. Therefore, in a crowded environment, the leading end of a long chain can be easily trapped between obstacles.

The formation of the spiral-coil is a peculiar example of such trappings. As discussed in Section~\ref{sec:sub-bending}, once the leading end of a long chain is forced to turn into a tighter conformation than the circle $\Pi$, it often slides along the inner side of the chain. This results in smaller free space in the middle of the conformation, and gradually the leading end keeps moving further into the center. Eventually the chain is self-packed as if it were coiled by a cable coiler.

This ordered conformation is very different from a collapsed polymer chain, say, in poor solvents, because the latter is disorderedly folded into a compact conformation. However, the spiral-coil is not always perfectly formed such that we could not simply use the Archimedean spiral to identify it. The long chain is also an obstacle for the short chains. As a result, at a high $C$ the long chain is often accompanied by a number of short chains, moving in the same direction with the local segment of the long chain, as shown in Fig.~\ref{fig:phase_diagram}(a).

In Section~\ref{sec:sub-bending} we also have discussed that the energetic cost of bending a long chain can be too high for spontaneously forming a spiral-coil if the persistence length is long. For these cases, the spiral conformation is maintained due to the excluded volume interaction from the short chains moving together with the rotating spiral, especially from those outsides the most outer thread. The clusters formed in the simulation are usually locked by the splay or wedge alignment of short chains. Such formations generate forces inwards to compensate the energetic penalty of bending and to prevent the folded long chain from unfolding. In addition these bead-forming chains effectively have higher friction than smooth bending rods when they are in contact due to the locked-in beads. As a result, a well-formed spiral-coil in the simulation rarely unfolds unless it is intervened by the collisions from another cluster that removes the constraints. Therefore, the properties of clusters, such as the size distribution and the number fluctuation, is crucial for the formation and the persistence of the spiral-coil. And these properties, as demonstrated in the simulation results, depend on the bending modulus of the self-propelled chain.

\section{\label{sec:conclusions}Conclusions}
In this article we have studied the effects of bending modulus and number density on the collective behavior of self-propelled chains and the formation of the spiral-coil. Our results show that, other than the number density, the bending modulus also plays an important role in the cluster formation and the giant number fluctuation of self-propelled chains. We find that the properties of clusters and the bending modulus essentially determine whether the self-propelled long chain mixed with the short chains folds into the spiral-coil. In particular, we also find that at a high number density of the background short chains, the probability of forming the spiral-coil is similar between the flexible and semiflexible long chains; both exhibit similar conformation in spite of the different bending moduli. The results support the previous speculation on the formation of the spiral-coil in the swarm of \emph{V. alginolyticus}. They also exemplify that it may be inadequate to derive the mechanical properties of a biopolymer in an active system simply by measuring its bending, as shown by the example of microtubules in Ref.~\cite{Brangwynne2007}.

As far as the generic properties of active matter are concerned, our simulation shows some unexpected results that require further investigation in order to understand their sources. First, the number fluctuation for the flexible chains shown in Fig.~\ref{fig:num_fluc} does not follow a simple power-law as those semiflexible cases. We are not aware of any theory to explain this behavior. Second, even though we have derived a simple theory to explain the smaller size of clusters among flexible chains, it does not explicitly predict the curves shown in Fig.~\ref{fig:CSD}. In the meantime, the mutual locking of chains in a cluster due to the bead-chain construct might be effectively an attraction between chains that have been overlooked in this article. Particularly at a high area fraction, the inward forces to keep a cluster from falling apart was enhanced in the simulation. Therefore, this could explain why the results presented in this article are different from those of the experiment of \emph{M. xanthus} reported in Ref.~\cite{Peruani2012}.

\section{Acknowledgements}\label{sec:Acknowledgement}
This work is financially supported by the Ministry of Science and Technology, Republic of China under contract No.~MOST-106-2112-M-008-023 and MOST-107-2112-M-008-025-MY3.
\bibliography{references}

\end{document}